\pdfoutput=1
\documentclass[aps,prl,reprint,groupedaddress,nofootinbib]{revtex4-1}
\usepackage[]{natbib}
\usepackage{amsmath, amssymb, amsthm, url}
\usepackage{graphicx}
\usepackage{xcolor}  
\usepackage[normalem]{ulem} 
\usepackage{dcolumn}
\newcolumntype{d}[1]{D{.}{.}{#1}}

\graphicspath{{Figures/}}

\DeclareMathOperator*{\argmin}{arg\,min}

\begin{document}


\title{Hybrid statistical and mechanistic mathematical model guides mobile health intervention for chronic pain}



\author{Sara M.~Clifton}
\email[E-mail me at: ]{sclifton@u.northwestern.edu}
\affiliation{Department of Engineering Sciences and Applied Mathematics, Northwestern University, Evanston, IL 60208, USA}

\author{Chaeryon Kang}
\affiliation{Department of Biostatistics, University of Pittsburgh, Pittsburgh, PA 15261, USA}

\author{Jingyi Jessica Li}
\affiliation{Department of Statistics, University of California Los Angeles, Los Angeles, CA 90095, USA}

\author{Qi Long}
\affiliation{Department of Biostatistics and Epidemiology, University of Pennsylvania, Philadelphia, PA 19104, USA}

\author{Nirmish Shah}
\affiliation{Department of Medicine, Duke University, Durham, NC 27710, USA}

\author{Daniel M.~Abrams}
\affiliation{Department of Engineering Sciences and Applied Mathematics, Northwestern University, Evanston, IL 60208, USA}
\affiliation{Northwestern Institute for Complex Systems, Northwestern University, Evanston, IL 60208, USA}
\affiliation{Department of Physics and Astronomy, Northwestern University, Evanston, IL 60208, USA}


\date{\today}

\begin{abstract}
Nearly a quarter of visits to the Emergency Department are for conditions that could have been managed via outpatient treatment; improvements that allow patients to quickly recognize and receive appropriate treatment are crucial. The growing popularity of mobile technology creates new opportunities for real-time adaptive medical intervention, and the simultaneous growth of ``big data'' sources allows for preparation of personalized recommendations. Here we focus on the reduction of chronic suffering in the sickle cell disease community. Sickle cell disease is a chronic blood disorder in which pain is the most frequent complication. There currently is no standard algorithm or analytical method for real-time adaptive treatment recommendations for pain.  Furthermore, current state-of-the-art methods have difficulty in handling continuous-time decision optimization using big data. Facing these challenges, in this study we aim to develop new mathematical tools for incorporating mobile technology into personalized treatment plans for pain. We present a new hybrid model for the dynamics of subjective pain that consists of a dynamical systems approach using differential equations to predict future pain levels, as well as a statistical approach tying system parameters to patient data (both personal characteristics and medication response history). Pilot testing of our approach suggests that it has significant potential to predict pain dynamics given patients' reported pain levels and medication usages. With more abundant data, our hybrid approach should allow physicians to make personalized, data driven recommendations for treating chronic pain.
\end{abstract}

\maketitle

\

\section{Introduction} \label{sec:intro}

In the fields of physics, chemistry, and engineering, models are often derived from mechanistic fundamental laws expressed in the form of differential equations.  Resulting ``dynamical systems'' models can be used both to gain intuition into the expected behavior of the system, and to make specific predictions about results of experiments (e.g., see \cite{strogatz2014nonlinear}). In fields such as social sciences, bioinformatics, and medicine, models are often constructed from data via statistical inference, without direct derivation from fundamental principles (e.g. \cite{freedman2009statistical}).  The mechanistic and statistical approaches to mathematical modeling have different advantages.  The former allows prior knowledge to be introduced and validated or rejected based on the success of the model.  The latter requires almost no a-priori information about how the system is expected to behave.

In this paper, we present a hybrid approach to mathematical modeling that incorporates both mechanistic and statistical elements, with the goal of gaining a deeper understanding of the human experience of subjective pain.  Specifically, we hope to predict how patient-reported pain levels vary over time based on medication dosage information and other patient characteristics.
\begin{figure*}[tb!]
    \centering
    \includegraphics[width=0.85\textwidth]{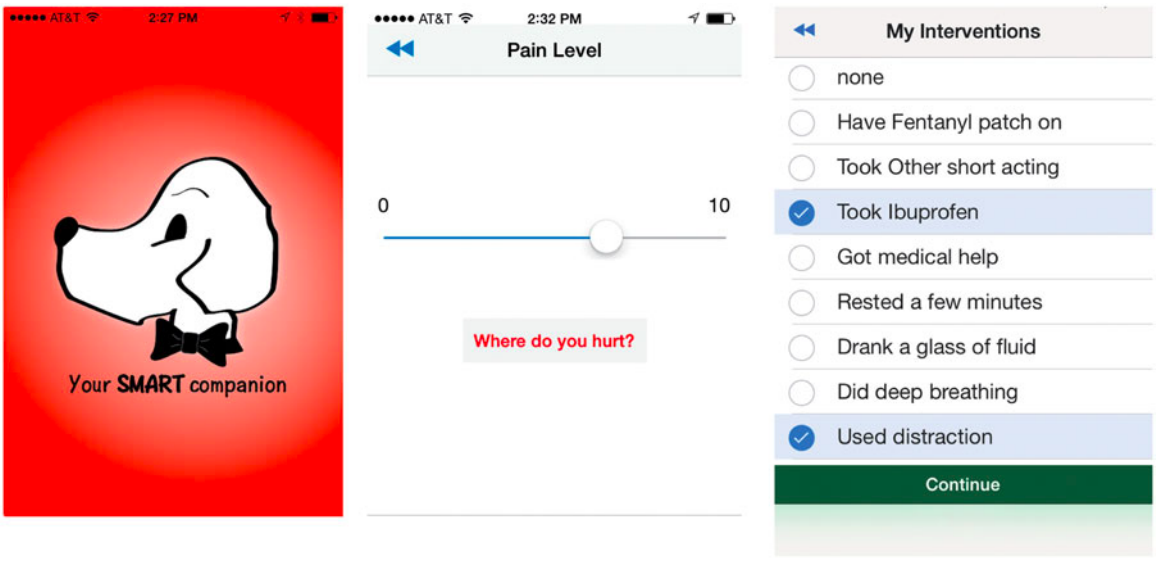}
    \caption[Sample images of SMART app for smartphone devices.]{\textbf{Smartphone app.} Sample images of SMART app for iPhone/Android smartphone devices.}
    \label{fig:screencaps}
\end{figure*}

\subsection{Application to pain}
Sickle cell disease (SCD) is a chronic illness associated with frequent medical complications and hospitalizations. Approximately 90\% of acute care visits are for pain events, and 30-day hospital re-utilization rates are alarmingly high \cite{platt1991pain}. While factors influencing these high re-utilization rates are poorly understood, close follow-up and continued use of pain medications has been shown to decrease re-hospitalization rates. Mobile technology has become an integral part of health care management, and our recently self-developed mobile application (Sickle cell Mobile Application to Record symptoms via Technology, or SMART app---see Figure \ref{fig:screencaps}) for SCD assists with documentation and intervention of pain.

Pain in particular is difficult to quantify and has never before been monitored at the temporal scale we report here across so many patients. It is known that subjective pain, though indeed subjective, is correlated with objective measurable stimuli qualities in experiments (see, e.g., \cite{granovsky2008objective, hughes2002assessment, stevens1958scale}).  Thus there is reason to believe that subjective pain may follow understandable dynamics in time, especially when mitigated by opioid or non-opioid drugs.  Our approach to the problem is motivated by the hope that a reasonable model for pain dynamics will yield some level of predictive power, despite the clear expectation that there will also be significant noise within and across patients.  We can attribute the stochastic variation to sources like patient mood, temporal changes in patient state, weather, etc.  In contrast, we hope that patient attributes like age, gender, SCD disease type, etc.~will remain roughly constant on the time scale of the experiment and allow us to explore possible correlation of these attributes with model parameters.

\subsection{Data source: mobile health application}
We seek to understand the temporal dynamics of chronic pain as experienced by SCD patients.  To that end, we have developed a mobile phone application that allows patients to record medication usage and subjective pain levels (measured on a 0-10 scale) in real time \cite{jonassaint2015usability, shah2014patients}.

\begin{figure}[tb!]
    \centering
    \includegraphics[width=\columnwidth]{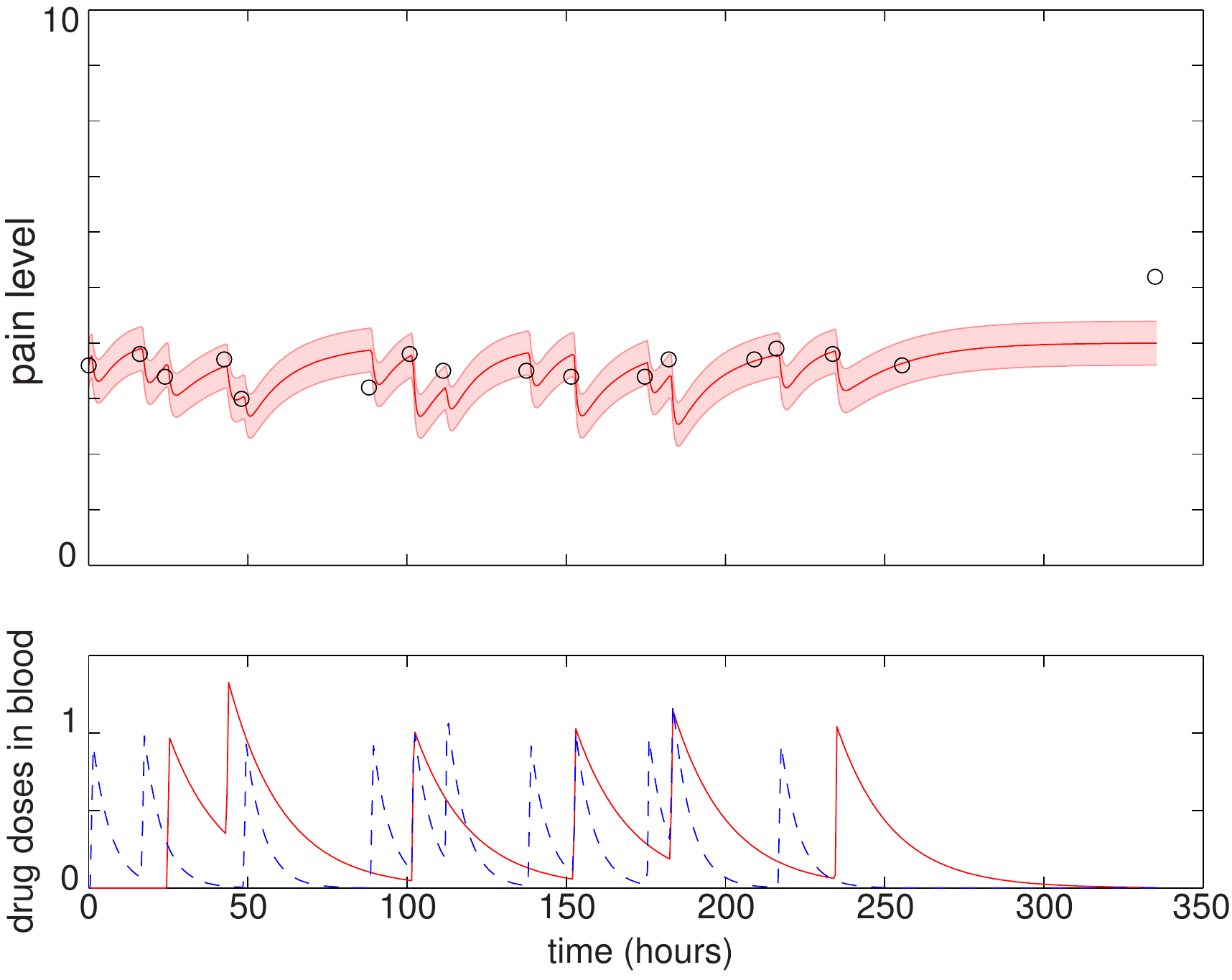}
    \caption[Sample pain and medication data from a single patient.]{\textbf{Sample pain and medication data from a single patient.} Upper panel: patient reported pain (black circles) and model fit (red solid line); red shading indicates model fit plus/minus one standard deviation. Lower panel: long-acting methadone (red solid line) and short-acting oxycodone (blue dashed line) medication concentrations in patient bloodstream as inferred from medication usage reported via the SMART application.}
    \label{fig:sampledata}
\end{figure}

Figure \ref{fig:screencaps} shows several images of the app interface, while Figure \ref{fig:sampledata} shows a typical data set resulting from a single patient's use of the app over the course of several weeks.

\section{Materials and Methods} \label{sec:methods}


\subsection{Data}
As of October 2016, data were available from 47 patients using the SMART app.  Data sets from 8 of those patients were excluded because of excessive sparsity based on the following criteria:(1) total number of reports $ \leq 5$; or (2) pain reports never exceeded zero during the period under consideration. See Table \ref{table:demo3} for demographic details of included patients. We denote the sample size $n=39$.

\begin{center}
  \begin{table*} \small
  \scriptsize

  \begin{tabular}{ l  r   d{7.1} c }
    Demographic characteristics       &     &     & \\
   \hline
                                          & N   & (\%)&\\
    \cline{2-3}
     Institution \hspace{0.1cm}         &      &   & \\
                  \hspace{4.7 cm} A     & 14   &( 35.9 ) &\\
                  \hspace{4.7 cm} B     & 17    &( 43.6 ) &\\
                  \hspace{4.7 cm} C     & 8    &( 20.5 ) &\\
                                        &      &         &\\
    Gender                                 &      &       &  \\
                 \hspace{4.7cm} Male    &  16  & ( 41.0 )& \\
                 \hspace{4.7cm} Female  &  23  & ( 59.0 ) &\\
    Age at baseline (years)             &      &          &\\
                 \hspace{4.7 cm} 18-34  &  24  & ( 61.5 ) &\\
                 \hspace{4.7 cm} $>34$  &  15  & ( 38.5 ) &\\
    SCD disease type                     &      &       &\\
                 \hspace{4.7 cm}Hemoglobin SC      &  8   & ( 20.5 )& \\
                 \hspace{4.7 cm}Hemoglobin SS     &  22  & ( 56.4 ) &\\
                 \hspace{4.7 cm}Hemoglobin SB$+$ (Beta) Thalassemia &   5  & ( 12.8 )  &\\
                 \hspace{4.7 cm}Beta-Zero Thalassemia    &   3  & ( 7.7 ) &\\
                 \hspace{4.7 cm}SO$-$Ara   &   1  & ( 2.6 ) &\\
    Hydroxyurea user                       &   27 & ( 69.2 ) &\\
    Folic acid vitamin user                &   26 & ( 66.7 ) &\\
    Long-acting opioid user            &   29 & ( 74.4 ) &\\
    Short-acting opioid user           &   35 &  ( 89.7 ) &\\
    Non-opioid user             &   29 &  ( 74.4 )&\\

                                        &      &          &\\     
                                        & Mean & \text{SD}  & (Min, Max) \\
    \cline{2-4}
    Number of pain reports                     &67.2 &  60.4 & ( 9.0, 257.0 )   \\     
    Days of pain reports                      & 164.6  & 109.6  & ( 10.3, 435.1 ) \\
    Within-patient average VAS score     & 4.7 &   2.1 & ( 0.3, 9.4 ) \\  
    
                                        &      &          &\\     
                                        & Mean & \text{SD}  & (Min, Max) \\
    \cline{2-4}
    Number of pain reports (first 2 weeks)                         & 13.2 &  9.6 & ( 2.0, 45.0 )   \\     
    Number of long-acting opioid doses (first 2 weeks)             & 6.0  & 8.4  & ( 0.0, 35.0 ) \\
    Number of short-acting opioid doses (first 2 weeks)            & 7.2  & 7.5  & ( 0.0, 35.0 ) \\
    Number of non-opioid doses (first 2 weeks)                     & 2.1  & 3.1  & ( 0.0, 12.0 ) \\

    \hline
  \end{tabular}
  \caption{Patient demographic information and the number of pain reports supplied by patients across entire study.}
  \label{table:demo3}
  \end{table*}
\end{center}

\subsection{Predictive model}

In order to develop a hybrid model that incorporates both a mechanistic a-priori knowledge-driven component and a statistical data-driven component, we divide tasks into two disjoint sets that fit these two categories; see the Discussion section for more context.  We begin with a ``dynamical systems'' model for subjective pain motivated by the hypothesis that human sensory systems function on a roughly ``return to setpoint'' basis \cite{mcruer1974mathematical, fors1988ability, britton1995mathematical, stepan2009delay}.  Any model of human pain response, however, will inevitably require specification of a variety of parameters determining the time scale(s) and degree of severity of the response.  The statistical modeling tasks employ patient data to infer parameters (1) from patient characteristics and population distributions and (2) from patient-specific pain and medication response history.

\begin{figure*}[ht]
  \centering
  \includegraphics[width=0.8\textwidth]{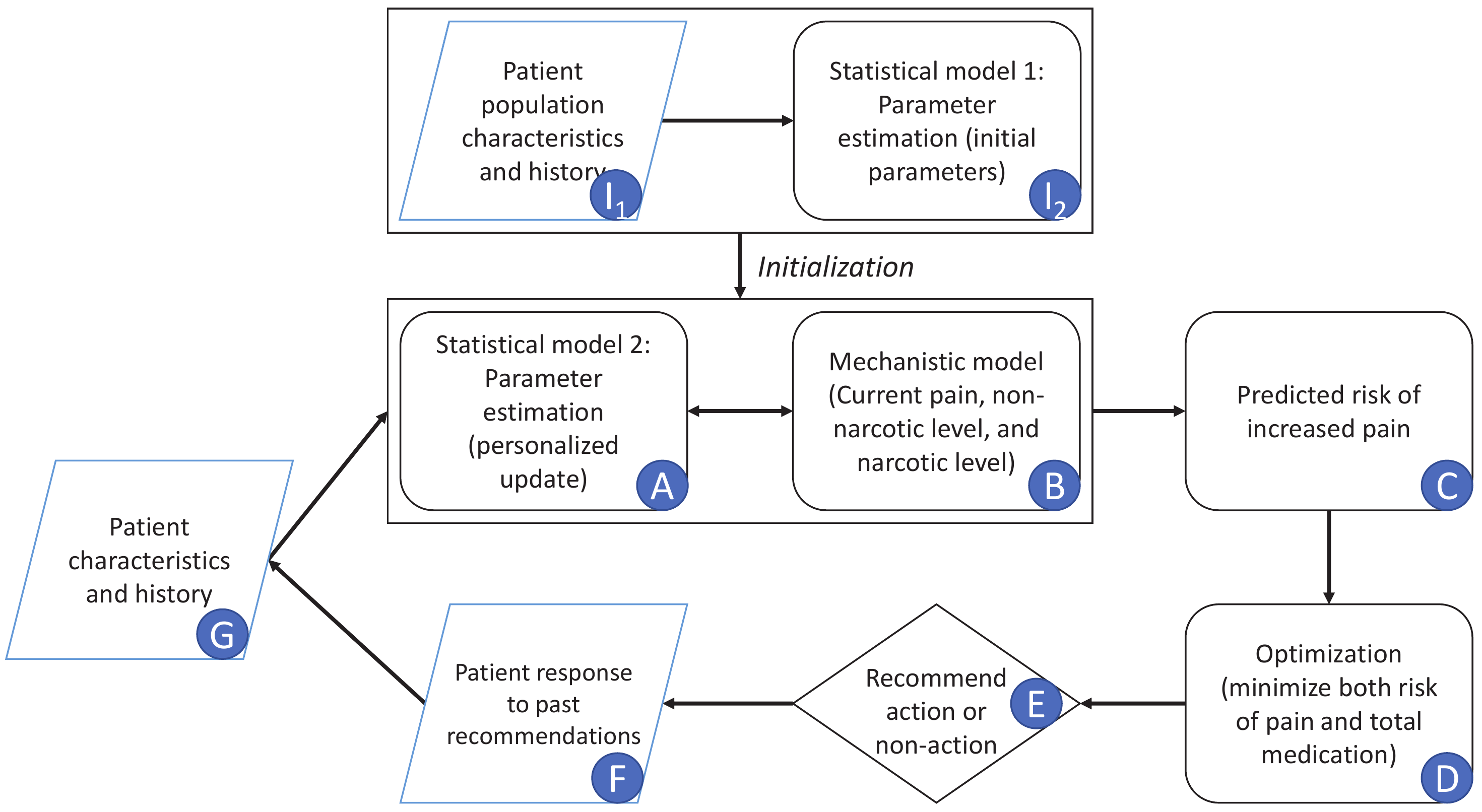}
  \caption[Schematic flowchart for hybrid model framework.]{\textbf{Schematic flowchart showing model framework.} Rounded rectangles represent modeling or computation steps, rhombuses represent data inputs or outputs, and diamond represents decision step.  Items $\textrm{I}_1$ and $\textrm{I}_2$ are only necessary for initialization of the model. Items $\textrm{A}$ through $\textrm{E}$ are the focus of this paper.}
  \label{fig:flowchart}
\end{figure*}

To make this more concrete, in Figure \ref{fig:flowchart} we present a flow chart summarizing our approach to the hybrid modeling problem.  Steps $\textrm{I}_2$ and A comprise the statistical modeling component; steps B and C comprise the mechanistic modeling component.  A further optimization step D builds on the predictions of the hybrid model to allow for a balance between competing demands of pain reduction and medication usage minimization. This paper details steps $\textrm{I}_1$, $\textrm{I}_2$, A and B. We leave the remaining steps for future work.

\subsubsection{Mechanistic component (for every patient)}
We propose and evaluate two related mechanistic models based on a set of coupled ordinary differential equations (ODEs), either (a) deterministic or (b) stochastic.  The stochastic differential equation (SDE) model comprises a Langevin equation, which can be converted into a Fokker-Planck partial differential equation (PDE) for the evolution of the probability distribution for pain $\rho(P,t)$ \cite{gardiner1985handbook}.  This allows for prediction of both the expected pain level for a patient at any point in the future \textit{and} an assessment of the confidence in (and a confidence interval for) that prediction.  

Mathematically, the deterministic mechanistic model we propose is the following, for a single patient:
\begin{equation}  \begin{aligned}
\frac{\mathrm{d} P}{\mathrm{d} t} &= -(k_0 +k_1 D_1+k_2 D_2 + k_3 D_3) P + k_0 u \\
\frac{\mathrm{d} D_1}{\mathrm{d} t} &= -k_{D_1} D_1 + \sum_{j=1}^{N_1} \delta(t-\tau_{1,j}) \\
\frac{\mathrm{d} D_2}{\mathrm{d} t} &= -k_{D_2} D_2 + \sum_{j=1}^{N_2} \delta(t-\tau_{2,j}) \\
\frac{\mathrm{d} D_3}{\mathrm{d} t} &= -k_{D_3} D_3 + \sum_{j=1}^{N_3} \delta(t-\tau_{3,j})\,,
\label{eq:painODE}
\end{aligned} \end{equation}
where $P$ is the patient pain level (on a scale of 1--10), $k_0$ is the pain relaxation rate without drugs, $k_i$ is the marginal effect on the pain relaxation rate due to drug $i$ ($i=1,2,3$), $u$ is the unmitigated pain level (i.e. without drug intervention), $D_i$ is the amount of standard drug $i$ doses within the patient, $k_{D_i}$ is the elimination rate of drug $i$ within the patient, $\{\tau_{i,j}\}_{j=1}^{N_i}$ are the drug $i$ dosage times, and $N_i$ is the number of doses of drug $i$ taken. $\delta$ represents the Dirac delta function.  Note that the parameters and variables will in general need to be indexed with distinct values for each patient in a population, though we omit those indices here for clarity. For convenience and clarity we also here omit ``hatted'' notation (e.g., $\hat{P}(t)$) sometimes used for model predictions.

Tables \ref{table:vars} and \ref{table:params} summarize the meanings of model variables and parameters, respectively.

\begin{center}
  \begin{table}[ht] \small
  \centering
  \begin{tabular}{ c  p{4.5cm}  l }
    \hline
    Variable & Meaning & Units \\
    \hline 
    $P(t)$ & Instantaneous pain level on 0--10 scale & [pain]  \\
    
    $D_1(t)$ & Concentration of drug 1 (long-acting opioid) in the body & [standard doses] \\
    
    $D_2(t)$ & Concentration of drug 2 (short-acting opioid) in the body & [standard doses] \\
    
    $D_3(t)$ & Concentration of drug 3 (non-opioid) in the body & [standard doses] \\
    
    $\rho(P,t)$ & Instantaneous probability distribution of pain level $P$ & [probability]  \\
    \hline
  \end{tabular}
  \caption{Variables in mechanistic models.}
  \label{table:vars}
  \end{table}
\end{center}

\begin{center}
  \begin{table*}[ht] \small
  \centering
  \begin{tabular}{ c  l  l }
    \hline
    Parameter & Meaning & Units \\
    \hline 
    $u$ & unmitigated pain level & [pain] \\
    
    $k_0$ & rate of decrease of pain in the absence of drugs or acute sources of pain & [$T^{-1}$] \\
    
    $k_1$ & effect of drug 1 (long-acting opioid) on pain relaxation rate & [$T^{-1}$] \\
    
    $k_2$ & effect of drug 2 (short-acting opiod) on pain relaxation rate & [$T^{-1}$] \\
    
     $k_3$ & effect of drug 3 (non-opioid) on pain relaxation rate & [$T^{-1}$] \\
    
    $k_{D_1}$ & rate of decay of drug 1 (long-acting opioid) in body due to metabolism  & [$T^{-1}$] \\
    
    $k_{D_2}$ & rate of decay of drug 2 (short-acting opioid) in body due to metabolism  & [$T^{-1}$]  \\
    
    $k_{D_3}$ & rate of decay of drug 3 (non-opioid) in body due to metabolism  & [$T^{-1}$]  \\
    
    $\varepsilon$ & amplitude of intrinsic variability in human subjective pain reports &  $[T^{1/2}]$ \\
    
    $N_i$ & number of standard drug $i$ doses taken & [count] \\ 
    
    $\{\tau_{i,j}\}$ & drug $i$ dose times (indexed by $j$) & $[T]$ \\ 
     \hline
  \end{tabular}
  \caption{Parameters in mechanistic models.}
  \label{table:params}
  \end{table*}
\end{center}

In this very simple model for pain dynamics \eqref{eq:painODE}, pain is expected to relax at rate $k_0$ to unmitigated level $u$ set by aggravating factors (like sickle cell disease) in the absence of intervention through opioids (drugs 1 and 2) or non-opioids (drug 3).  When drugs are present in the patient's body, pain drops at a faster rate and the short-term equilibrium pain level (not the unmitigated pain level $u$) is reduced. Note that we treat all parameters as constant over the time period of interest, which we take to be two weeks (based on clinical heuristic experience). 

In the model for drug concentrations, medication in the body is assumed to be metabolized at a constant rate.  Rates can be determined from existing substantiated pharmacokinetic models (e.g., \cite{yang2006,poulin2002}); Dirac delta function onset of medication serum concentration is a good approximation to the fast rise typical of the medications under consideration.  See Fig.~\ref{fig:sampledata} for a sample deterministic model output.

Note that we deliberately chose to employ an extremely simple conceptual model for pain dynamics. More sophisticated versions might be developed to incorporate higher order dynamics for $P$, or to include nonlinear or nonautonomous effects (e.g., allowing for explicit parameter variation with time of day or year), but currently available data are insufficient to constrain a model of greater complexity.

\begin{figure}[hbt]
  \centering
  \includegraphics[width=\columnwidth]{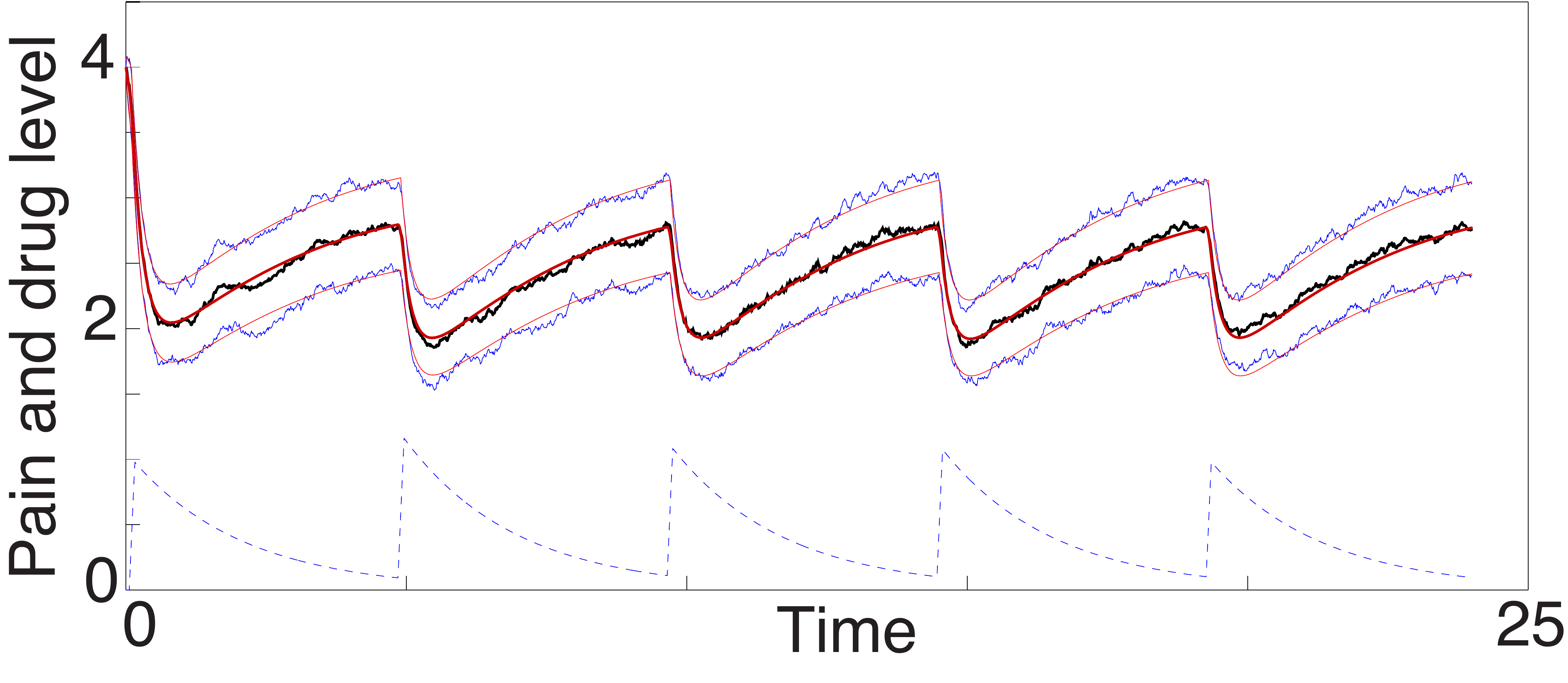}
  \caption[Sample output from stochastic differential equation model]{\textbf{Sample output from stochastic differential equation model \eqref{eq:langevin}.} Red thick line: theoretical mean pain; red thin lines: $\pm$ one theoretical standard deviation; black thick line: mean of pain distribution in ensemble of 100 stochastic simulations; blue thin lines: $\pm$ one standard deviation in ensemble of 100 stochastic simulations; blue dashed line: drug 1 dose in bloodstream. Spikes occur when patient takes recommended dosage.}
  \label{fig:sampleoutput_FP}
\end{figure}

The stochastic differential (Langevin) equation version of our mechanistic model is as follows:
 \begin{equation}\begin{aligned}
	 \mathrm{d} P &= -(k_0 + k_1 D_1 + k_2 D_1) P \mathrm{d}t + k_0 u (\mathrm{d} t + \varepsilon \mathrm{d}W) \\
	 \mathrm{d} D_1 &= \left(-k_{D1} D_1 + \sum_{j=1}^{N_1} \delta(t-\tau_{1,j}) \right) \mathrm{d} t, \\
	 \mathrm{d} D_2 &= \left(-k_{D2} D_2 + \sum_{j=1}^{N_2} \delta(t-\tau_{2,j}) \right) \mathrm{d} t \\
     \mathrm{d} D_3 &= \left(-k_{D3} D_3 + \sum_{j=1}^{N_3} \delta(t-\tau_{3,j}) \right) \mathrm{d} t\,,
	  \end{aligned}
	  \label{eq:langevin}
\end{equation}
where a hypothesis of uncorrelated additive white noise has been made. From this we derive the Fokker-Planck equation for the probability distribution of pain over time $\rho(P, t)$:
 \begin{equation} \begin{split}
	\frac{\partial \rho}{\partial t} = - \frac{\partial}{\partial P}\bigg[ \big(-(k_0 +k_1 D_1 +k_2 D_2 + k_3 D_3) P + \\ k_0 u\big) \rho(P,t) \bigg] + \frac{\partial^2}{\partial P^2}\bigg[ \frac{1}{2} (\varepsilon k_0 u)^2 \rho(P,t) \bigg]~.
	  \label{eq:fp}
\end{split}\end{equation}
Absent any pain medication, this Fokker-Planck equation implies the steady-state pain distribution
\begin{equation}
\rho^*(P) = \sqrt{\frac{1}{\pi k_0 u^2 \varepsilon^2 }} \exp\left[- \frac{(P-u)^2}{k_0 u^2 \varepsilon^2}\right],
\label{eq:steadystate}
\end{equation}
a Gaussian distribution with mean $u$ and standard deviation $ u \varepsilon \sqrt{k_0 / 2} $.  See Figure \ref{fig:sampleoutput_FP} for a sample stochastic model output. 

\subsubsection{Statistical component (for all patients)}
In order to account for the variation among patients and improve prediction of the unmitigated pain level, we associate patient characteristics and history with the unmitigated pain level $u$ (an $n$-dimensional vector with $u_j$ corresponding to the $j$-th patient's unmitigated pain level) using a linear model. 

Let $X$ be an $n \times p$ design matrix containing the covariates of patients (i.e., patient characteristics). We write $X=(X_1, \ldots, X_n)^T$, with $X_j$ corresponding to the $p$-dimensional covariates of patient $j$. Then we formulate the relationship between between $u$ and the $p$ predictors as:
\begin{eqnarray}
\label{eq:s4}
u= X \beta + \epsilon,
\end{eqnarray}
where $\beta$ is a $p$-dimensional coefficient vector, and $\epsilon$ is an $n$-dimensional vector of zero-mean random errors. When $p$ is small, the estimate for $\beta$ is obtained using the ordinary least squares procedure: $\displaystyle \hat{\beta}=\argmin_{\beta \in \mathbb{R}^p} \|u-X\beta\|_2^2$, where $\|\cdot\|_q$ denotes the $\ell_q$ norm.  Then the unmitigated pain level $u_j$ is updated by $u_j^{new}=X_j^T \hat{\beta}$, $j = 1,\ldots, n$. 

Since the unmitigated pain levels are not observable from patient pain reports, the initial $u_j$'s are independently sampled from a uniform distribution between $0$ and $10$, i.e. $u_j^{0} \sim U(0,10)$. After using $\{u_j^{0} \}$ as initial values to fit the mechanistic model, the resulting estimated $\{u_j\}$ will be updated by the linear model (\ref{eq:s4}) as $\{u_j^{new}\}$, which will then be used as initial values in the next round of fitting of the mechanistic model. See Section \ref{sec:modelfitting} for more detail on the hybridization of the statistical component with the mechanistic component.

Given a high-dimensional set of patient characteristics, we need to select a subset of patient characteristics that are significantly associated with $u$ by minimizing the penalized loss function. In this study, we select patient characteristics using the LASSO (Least Absolute Shrinkage and Selection Operator) \cite{tibshirani1996regression}, by minimizing the penalized loss function
 $\Gamma (\beta)=\| u- X\beta\|_2^2 + \lambda \|\beta\|_1$ with respect to $\beta$. The penalty parameter $\lambda$ is determined using 5-fold cross-validation. The selected $p$ features are then used to fit the linear model (\ref{eq:s4}) by ordinary least squares.

If time-varying unmitigated pain levels and time-varying covariates are present, the regression model \eqref{eq:s4} can be extended to the linear mixed model \cite{diggle2002analysis,fitzmaurice2012applied}: $u=X\beta + Z\delta + \epsilon,$ where $Z=(Z_1, \ldots, Z_n)^T$ is an $n \times r$ design matrix for $r$ random effect factors and
$\delta = (\delta_1, \ldots, \delta_r)^T$ is a vector of random effects. Patient characteristics can be selected by maximizing the penalized log-likelihood:  $\ell_{\mbox{pen}} (\beta, \delta)=\ell(\beta, \delta) - \lambda \|\beta\|_1$ \cite{groll2014variable}.  Such an extension of the proposed hybrid model to allow for time-varying unmitigated pain levels and covariates will be considered in a future study with more data available.

\subsection{Model fitting} \label{sec:modelfitting}

We fit our model to real patient data by minimizing the residual sum of squares between model predictions and patient reports provided within the first two weeks of reporting. We expect that the assumption of constant model parameters breaks down after approximately two weeks (clinical heuristics). Minimization over parameters $u$, $k_1, k_2$, and $k_3$ was done via the Nelder-Mead simplex algorithm \cite{nelder1965simplex}. Parameter $k_0$ was fixed at $2\ln(2) \approx 1.4$ corresponding to a pain equilibration half-life time scale of 30 minutes in the absence of medication.  If a patient did not take all three classes of drugs, the model and fitting only included the consumed drugs.

We initialize the parameter optimization in $n$ mechanistic models (one per patient) with random values during a first iteration, then we feed the optimization output into the statistical model (for all patients).  Once the statistical model is run, it results in a new set of parameter estimates that can then be employed as initial parameter seeds for a second round of optimization in $n$ mechanistic models (to minimize the residual sum of squares).  Proceeding iteratively in this fashion (see Fig.~\ref{fig:flowchart}), we find convergence to a consistent set of parameters for each patient (details below).

\subsection{Method verification}

Before applying our hybrid model to real-world patient data, we verify the soundness of the approach with synthetic data constructed to resemble real-world data, but generated by the model itself with high sampling frequency. The synthetic data used for verification of the method are generated directly from the mechanistic model with an assumed parameter set generated in the following way:  unmitigated pain $u = (\textrm{patient age})/10$, initial pain level $P(t=0) = u - 2$, and drug parameters $k_1,k_2,k_3 \sim N(0.75, 0.25)$.  Each patient reports pain every 1/2 hour for 336 hours (two weeks). At each report time, the probability of the patient taking a particular drug (among three drug classes) is 1/16; in other words, the patients took each drug on average every 8 hours. White noise of magnitude 1 is added to each pain report.

\begin{figure}[thb]
  \centering
  \includegraphics[width=\columnwidth]{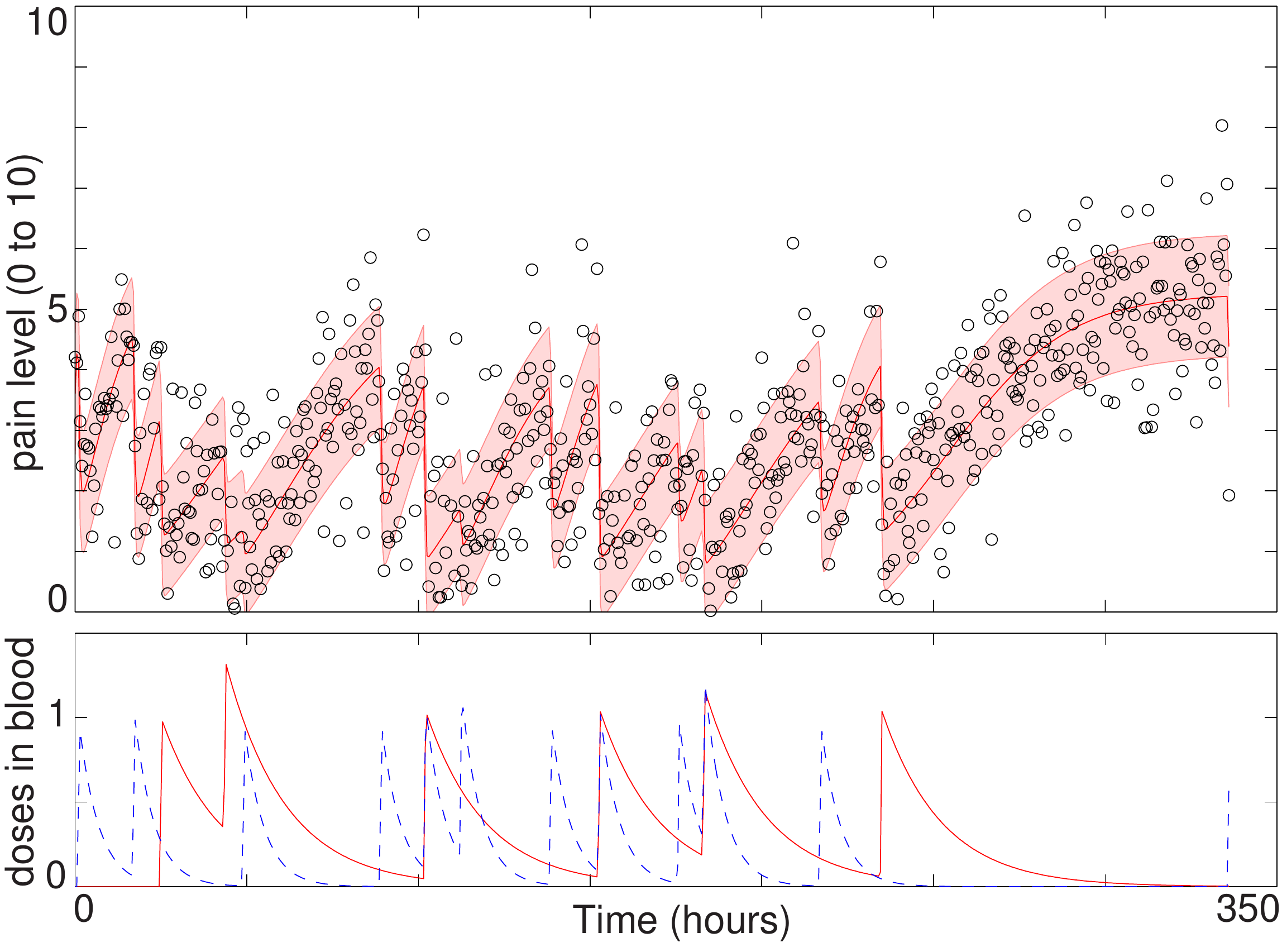}
  \caption[Model fitting demonstration for densely reported noisy synthetic data.]{\textbf{Model fitting demonstration for densely reported noisy synthetic data.} Upper panel: hypothetical densely-reported patient pain (black circles) and model fit (red solid line); red shading indicates model fit plus/minus one standard deviation. Lower panel: long-acting opioid (red solid line) and short-acting opioid (blue dashed line) medication concentration in patient bloodstream.}
  \label{fig:sample_modelfit_noise}
\end{figure}

As an illustration using \textit{real} patient drug times (specifically those of Patient A3), we create synthetic data generated using $u=5, k_1=3,$ and $k_2=2$: see Figure \ref{fig:sample_modelfit_noise}. When the initial parameter search is seeded with random parameter values, the mechanistic model fit can lead to convergence to either the ``true'' optimum $(5,3,2)$ or to other ``spurious'' optima with incorrect values of $u$, $k_1$, and $k_2$.  

In this illustrative example, the method converges to $u=5.01, k_1=3.19,$ and $k_2=1.84$. The relevant root-mean-square (RMS) error is 1.01; this is close to the lowest possible expected error given the unit magnitude noise added to the synthetic data. This numerical experiment shows that the mechanistic model fitting method can converge even in the presence of significant amounts of noise.  However, with \textit{only} the mechanistic model it can be quite difficult to find a good set of initial parameter seeds\footnote{The seeding problem becomes exponentially harder as the dimension of the parameter space increases.}: that is one motivation for introducing the statistical model.

\begin{figure}[thb]
  \centering
  \includegraphics[width=0.8\columnwidth]{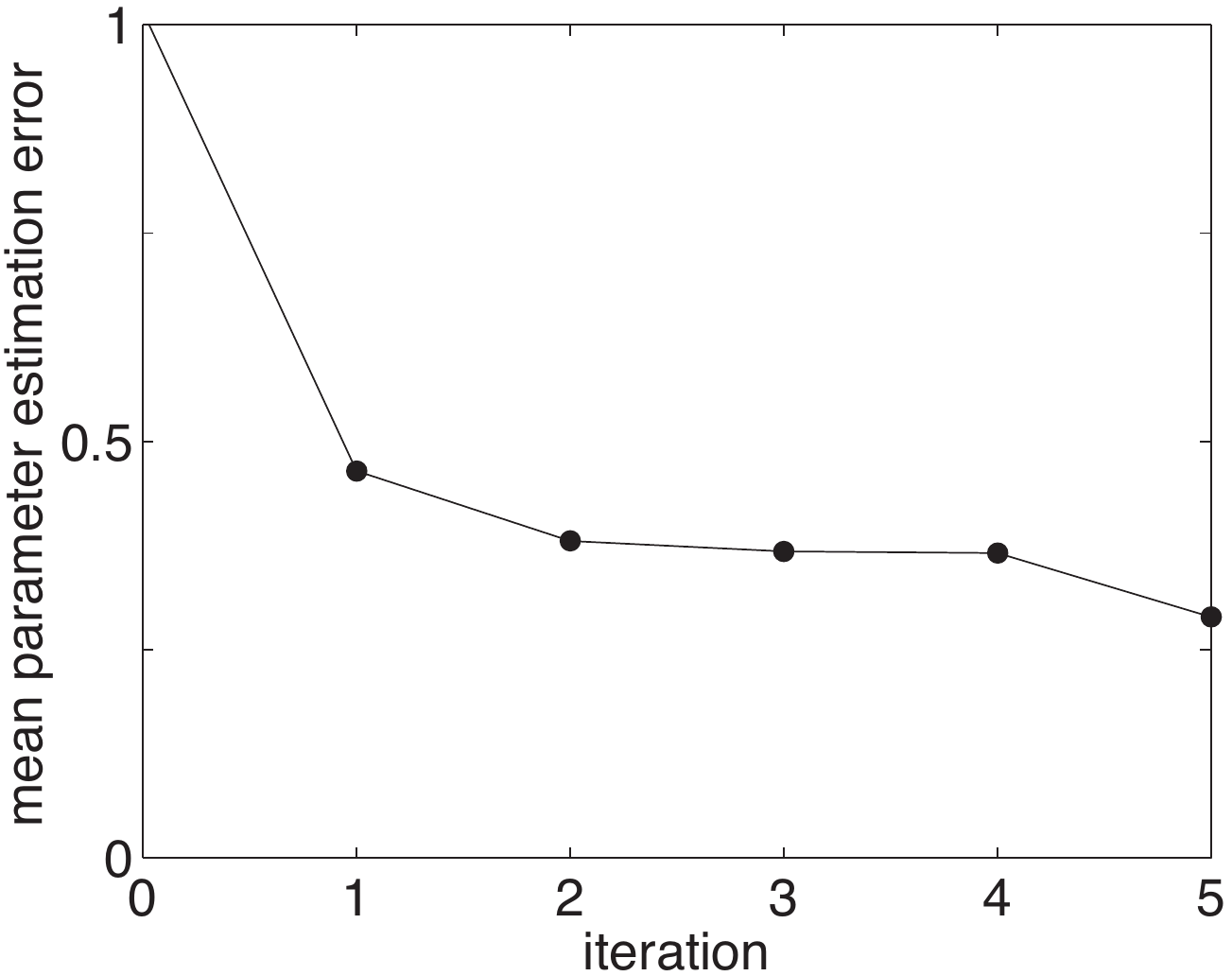}
  \caption[Hybrid model parameter fitting demonstration for ensemble of densely reported, noisy synthetic data.]{\textbf{Hybrid model fitting demonstration for ensemble of densely reported noisy synthetic data.} For an ensemble of 39 synthetic patient data sets, the average absolute error in $u$ gradually decreases. Iteration 0 indicates one fit to the mechanistic model alone. Subsequent iterations indicate the number of hybrid model (statistical + mechanistic) fits.}
  \label{fig:iter_error_u}
\end{figure}

\begin{figure}[thb]
  \centering
  \includegraphics[width=0.8\columnwidth]{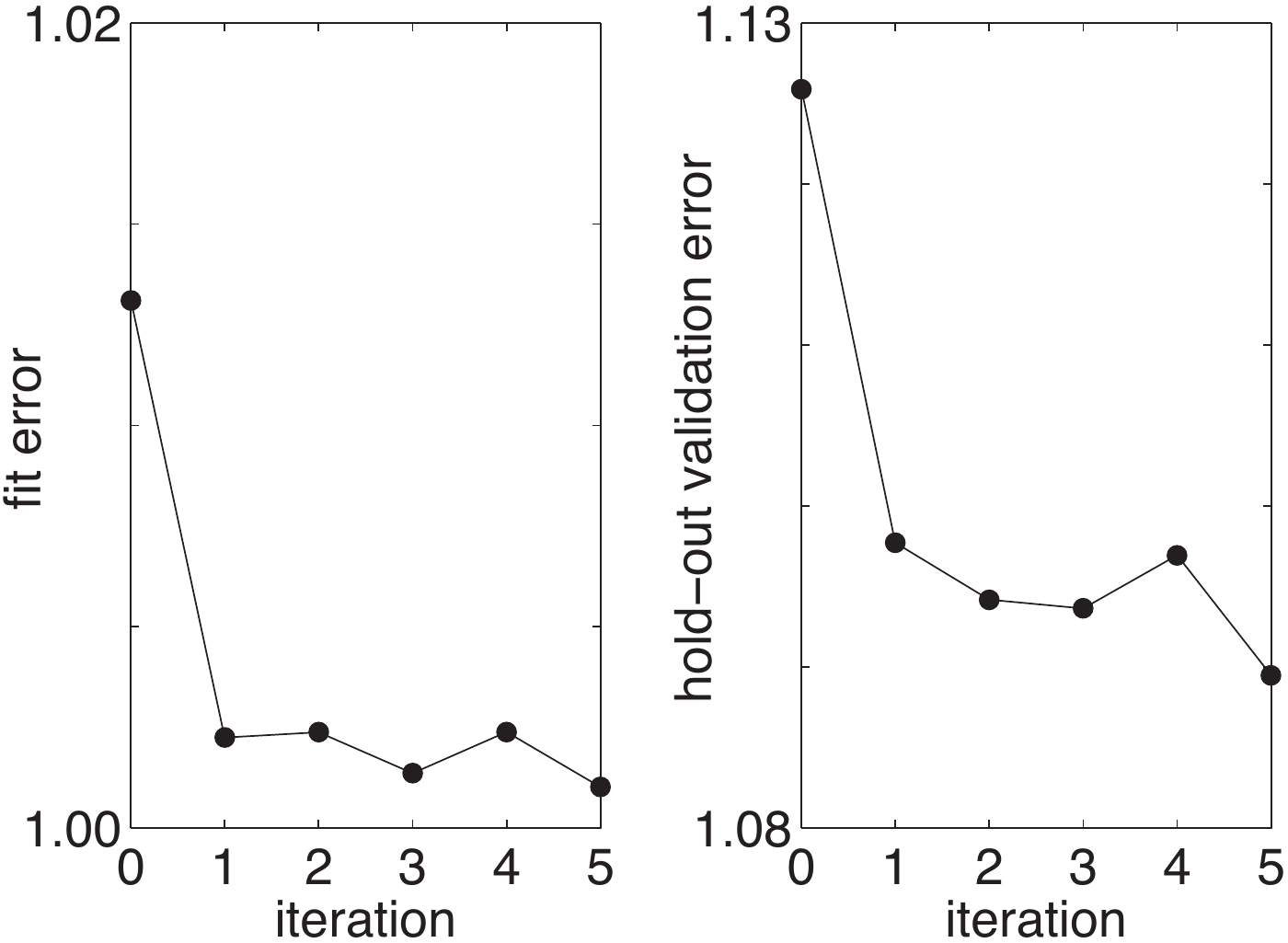}
  \caption[Validation of model using ensemble of densely reported, noisy synthetic data.]{\textbf{Hybrid model fitting demonstration for ensemble of densely reported noisy synthetic data.} For an ensemble of 39 synthetic patient data sets, the average root-mean-squared (RMS) error in patient pain levels gradually decreases. Iteration 0 indicates one fit to the mechanistic model alone. Subsequent iterations indicate the number of hybrid model (statistical+mechanistic) fits. Training error (or fit error) is on the left; test error (or validation error) is on the right. Due to the additive white noise of magnitude 1, the smallest testing or training error we could expect is 1.}
  \label{fig:iter_error_rms}
\end{figure}

To test our hybrid method using both the mechanistic model for fitting and the statistical model for parameter estimation, we create a synthetic patient database of 39 patients as described above.  We then iterate rounds of fitting between mechanistic and statistical models, starting with uniform random guesses for all patient parameters $(u, k_1, k_2, k_3)$.  Figure \ref{fig:iter_error_u} demonstrates how the parameter $u$ converges to a value with small error after just a few iterations steps, even in the presence of significant noise. In order to evaluate the performance of the model on new data, we use the hold-out validation method by splitting the dataset into a training set (first week) and a test set (second week). Model fit error and hold-out validation error, as well as other parameters values, converge similarly: see Figure \ref{fig:iter_error_rms}.

\section{Results}

\subsection{General results}
\begin{figure}[hbt]
  \centering
  \includegraphics[width=0.8\columnwidth]{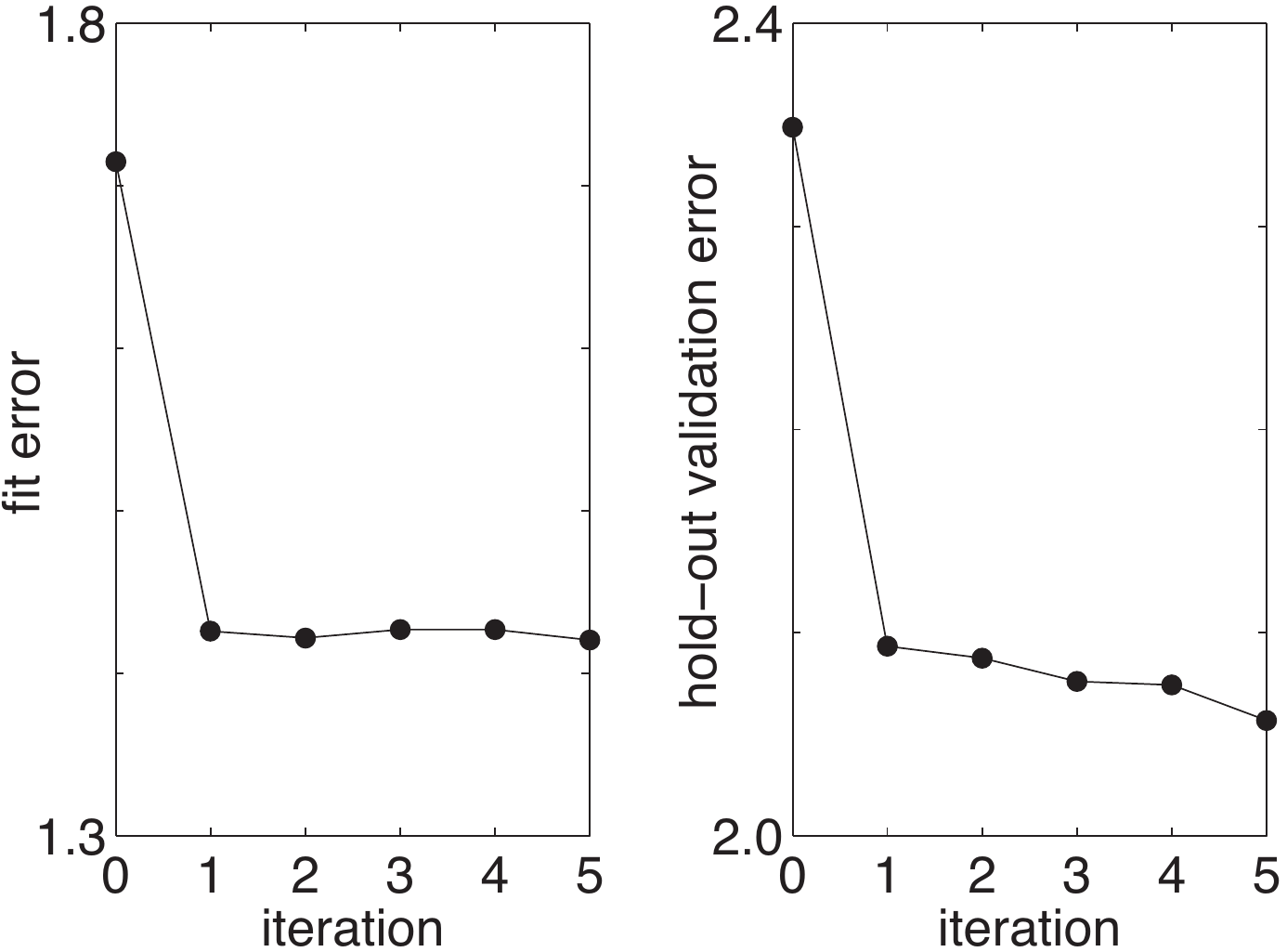}
  \caption[Hybrid model fitting on real patient data.]{\textbf{Hybrid model fitting on real patient data.} For an ensemble of 39 real patient data sets, the average root-mean-squared (RMS) error in patient pain levels gradually decreases. Iteration 0 indicates one fit to the mechanistic model alone. Subsequent iterations indicate the number of hybrid model (statistical+mechanistic) fits. Training error (or fit error) is on the left; test error (or validation error) is on the right.}
  \label{fig:iter_error_real_rms}
\end{figure}

One key result is that our model for chronic pain does indeed have some predictive value (see Fig.~\ref{fig:iter_error_real_rms}).  This is an improvement over the current state of the art, since no other predictive model exists of which we are aware. Furthermore, fitted parameter values correlate significantly with patient characteristics, suggesting that meaningful information is captured by this minimal plausible model. It may be possible to motivate new clinical insight on the basis of the observed correlations, perhaps leading to differential treatment of SCD sufferers with differing characteristics.

\subsection{Statistical results}
We use the following baseline patient characteristics to predict the unmitigated pain levels in the statistical modeling step: age, gender, SCD disease type, hydroxyurea use, folic acid vitamin use, long-acting opioid use, short-acting opioid use, and non-opioid use. We explore the marginal effects of these characteristics and their possible pairwise two-way interactions using the LASSO.
 The model \eqref{eq:s4}
can be extended to include time-varying covariates such as temperature, weather, patient's walking/social activities, and patient's mood at time $t$, once these data become available in a future study.

The statistical model that resulted from the LASSO variable selection is given by
\begin{align}
\label{eq:reg1}
\begin{split}
u = & \,\, \beta_0+\beta_1 \mathbf{1}\{\text{SCD disease type $=$ HgbSC}\} \\
     &+ \beta_2 \mathbf{1}\{\text{SCD disease type $=$ SB$+$Thal or SO-Ara}\} \\
 &+ \beta_3 (\text{age}-18) + \beta_4 \mathbf{1}\{\text{Hydroxyurea user}\} \\
 &+ \beta_5 \mathbf{1}\{\text{Non-opioid user } \}  \\
 &+ \beta_6 \mathbf{1}\{\text{SCD disease type $=$ SB$+$Thal or So-Ara}\} \\
 & \times (\text{age}-18) +\epsilon,
\end{split} \end{align}
where $\epsilon_j \sim N(0, \sigma^2)$, $j=1,\ldots,n$, and $\mathbf{1}\{\cdot\}$ is the indicator function.

Table \ref{table:PredUreg} summarizes the results from one round of fitting of the regression model \eqref{eq:reg1}. Adjusting for the effect of other terms in the regression model, SCD disease type of SB$+$Thal or So-Ara (with coefficient $\beta_2$), non-opioid use  (with coefficient $\beta_5$), and the interaction term between SCD disease type of SB$+$Thal or So-Ara and age  (with coefficient $\beta_6$) are important predictors of the unmitigated pain levels at the significant level of 0.05. Using non-opioid medication is associated with decreased unmitigated pain levels. Unmitigated pain levels increase with patients' age for SB$+$Thal or So-Ara patients. 

\begin{center}
  \begin{table*}[ht] \small
  \centering
  \begin{tabular}{ l r  r  r   r l}
    \hline
    Variable                      & \text{Estimate} & \text{Std Err} & \text{T-value} & \text{P-value} & \\
    \hline 
    Intercept               &  7.646  &  1.228  & 6.228 & 0.000  & *** \\
    HgbSC                   &  -1.566 &   0.890 & -1.761&  0.088 &  \\
    SB$+$Thal or So-Ara     &   -5.479&  2.332 & -2.349 & 0.025 & *  \\
    Age at baseline $-$18   &   0.001  &  0.034 &  0.290 &  0.773 &  \\
    Hydroxyurea user        &    -1.205  &  0.839 & -1.437 & 0.160 &   \\
    Non-opioid user     &   -2.523  &   0.842 & -2.995 & 0.005 & **   \\
    (SB$+$Thal or So-Ara) $\times$ (Age at baseline -18)
                            &  0.241  &  0.010 &  2.419 & 0.021 & *  \\
   \hline
   \end{tabular}
  \caption{Result of the prediction model of the unmitigated pain using the linear regression model. Significance $^{*} (p<0.05), ^{**} (p<0.01), ^{***} (p<0.001)$}
  \label{table:PredUreg}
  \end{table*}
\end{center}

\subsection{Mechanistic model validation}
\begin{figure}[htb!]
  \centering
    \includegraphics[width=\columnwidth]{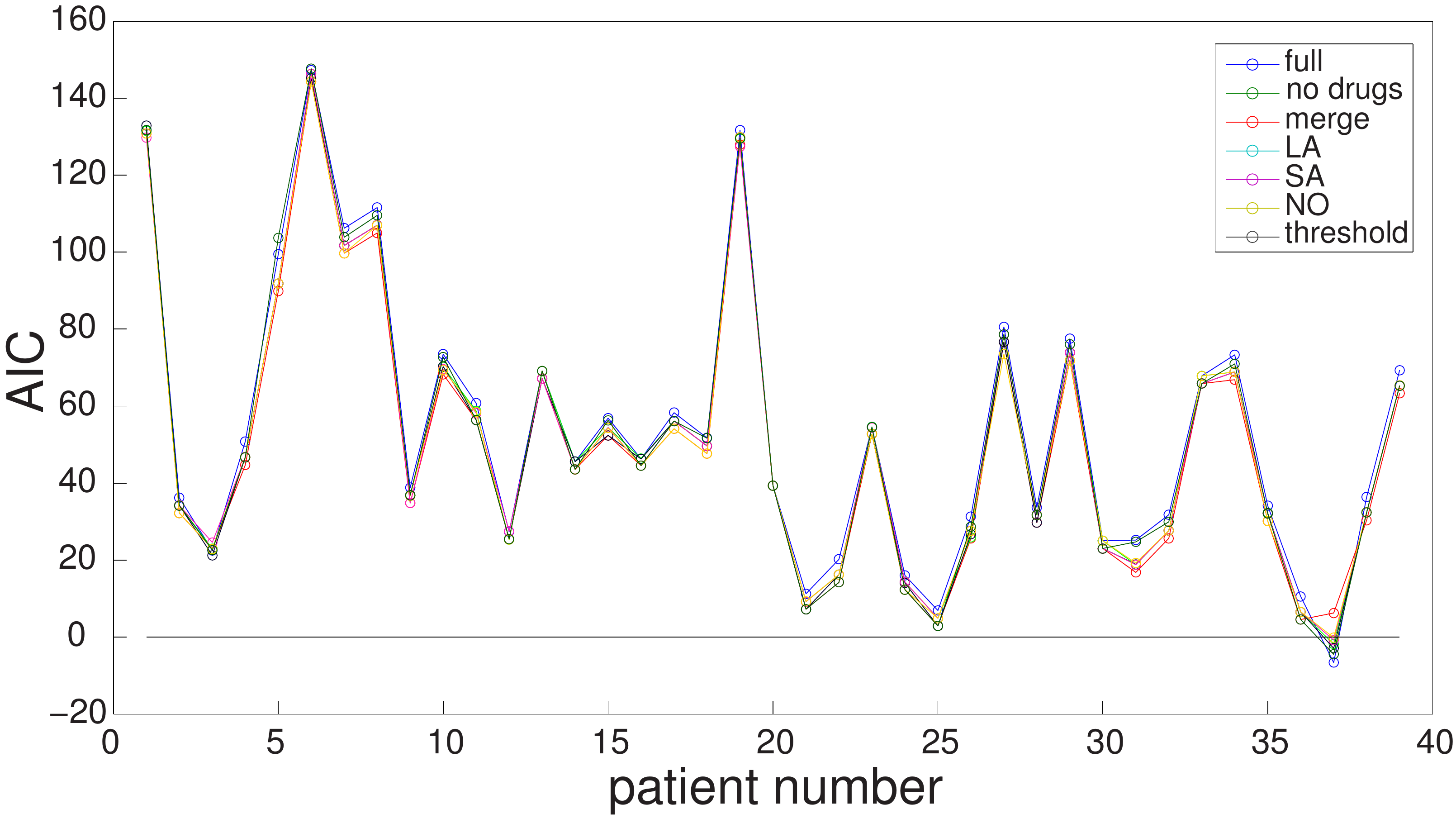}
     \caption[Akaike information criterion for alternative models.]{Akaike information criterion (AIC) for alternative models listed in Table \ref{table:models}. Most models perform equally well; among patients with differing model performance, there exists no clear `best' model for all patients.}
    \label{fig:aic}
\end{figure}

\begin{figure}[htb!]
  \centering
    \includegraphics[width=\columnwidth]{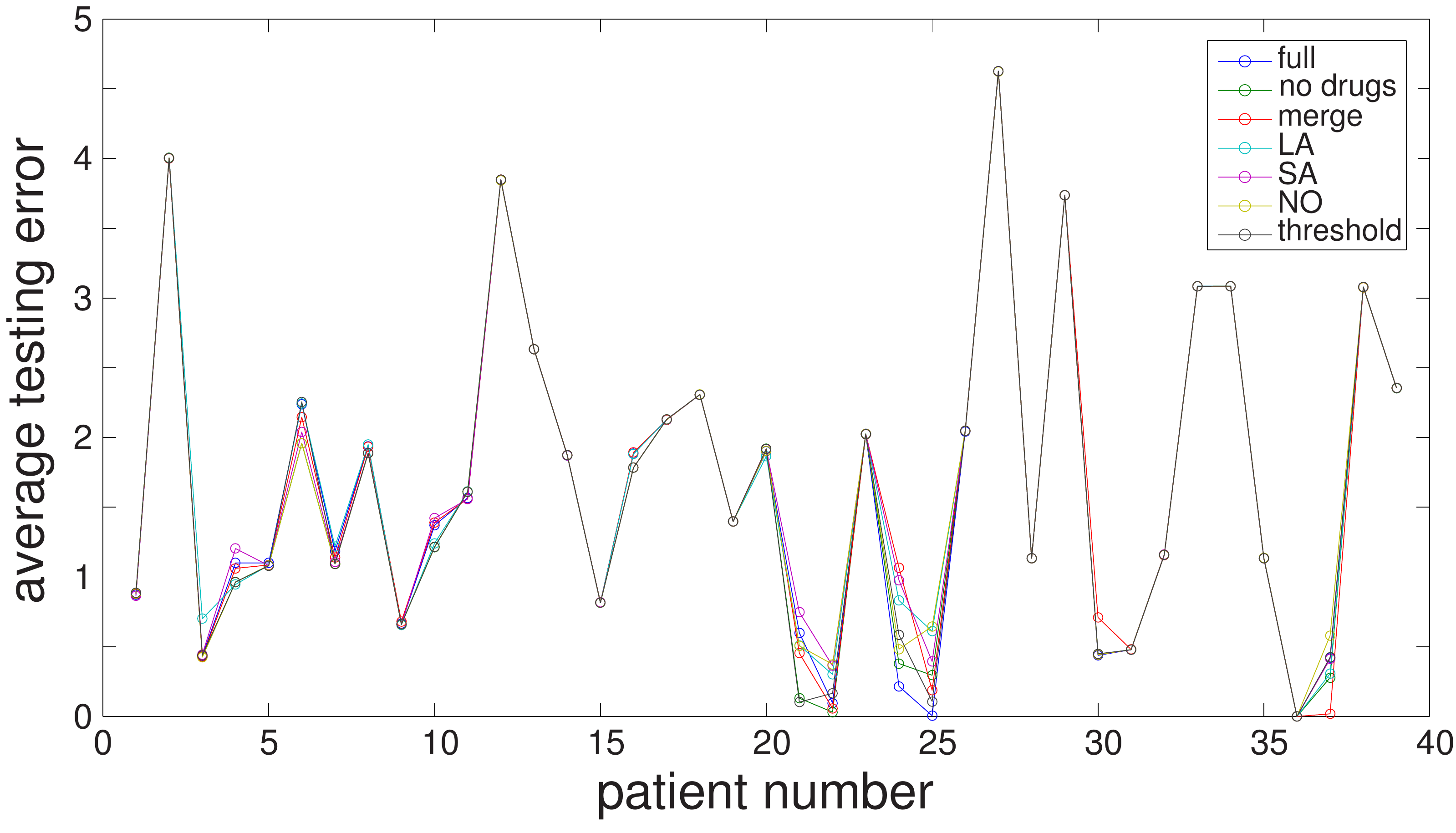}
     \caption[Two-fold cross validation testing error for alternative models.]{Two-fold cross validation testing error for alternative models listed in Table \ref{table:models}. For every patient, each model was independently fitted to the first and second half of the time series pain report data (training). Then the fitted models were used to test the other half of the data. This figure shows the average root-mean-square testing error for the two tests. Most models perform equally well; among patients with differing model performance, there exists no clear `best' model for all patients. Note that patient \#36 did not have enough data to fit any models, so zero error is misleading.}
    \label{fig:testerr}
\end{figure}

With such sparse data and up to four fitting parameters, one may worry that the model \eqref{eq:painODE} is being overfitted. To test this concern, we propose 6 related alternative models with fewer fitting parameters, and we compare cross-validation error and Akaike information criterion (AIC) among the models. See Table \ref{table:models} for model descriptions. Neither measure selected a best-fit model across all patients, but none of these simple models is overfitting the data. See Figure \ref{fig:aic} for AIC results and Figure \ref{fig:testerr} for cross validation results.

\begin{center}
  \begin{table*}[ht] \small
  \centering
  \begin{tabular}{ c  p{10cm}  l }
    \hline
    Model name & Description & Fitting parameters \\
    \hline 
    Full model & Include all drugs taken (model \eqref{eq:painODE}) & up to 4  \\
    
    No drugs & Include no drug dosing information & 1 \\
    
    Merge drugs & Combine all drugs into one drug class with same response & up to 2 \\
    
    LA only & Include only long-acting opioid doses & up to 2 \\
    
    SA only & Include only short-acting opioid doses & up to 2 \\
    
    NO only & Include only non-opioid doses & up to 2  \\
    
    Threshold & Include drug class only if drug is taken at least $n$ times* & up to 4  \\
    \hline
  \end{tabular}
  \caption[Mechanistic model variations.]{Mechanistic model variations. Fitting parameters include unmitigated pain level $u$ and drug response parameters $k_i$ for all drugs consumed. Therefore some patients have fewer fitting parameters than listed if they consumed fewer than three types of drugs. *In our tests, $n=5$ was the drug dose threshold.}
  \label{table:models}
  \end{table*}
\end{center}

\section{Biased pain reporting}
Perhaps the most significant limitation of our model lies in a potential bias in our data set. Patients typically report pain levels when taking medication, but many of them only take medication when pain levels rise. Thus we suspect a selection bias of unknown significance, causing higher pain levels to be reported at a disproportionately high rate. To test this concern, we compare the unbiased model \eqref{eq:painODE} with a similar model incorporating biased pain reporting.

Suppose the probability density function of pain at a particular time is a normal distribution with mean $\mu$ and variance $\sigma^2$: 
\begin{equation}
\rho (x \, | \, \mu, \sigma) = \frac{1}{\sqrt{2 \pi \sigma^2 }} \exp\left[- \frac{(x-\mu)^2}{2 \sigma^2}\right].
\label{eq:p}
\end{equation}
Integrating model \eqref{eq:painODE} gives the expected pain value $\mu$ at any point in time. 

\begin{figure}[htb!]
  \centering
    \includegraphics[width=\columnwidth]{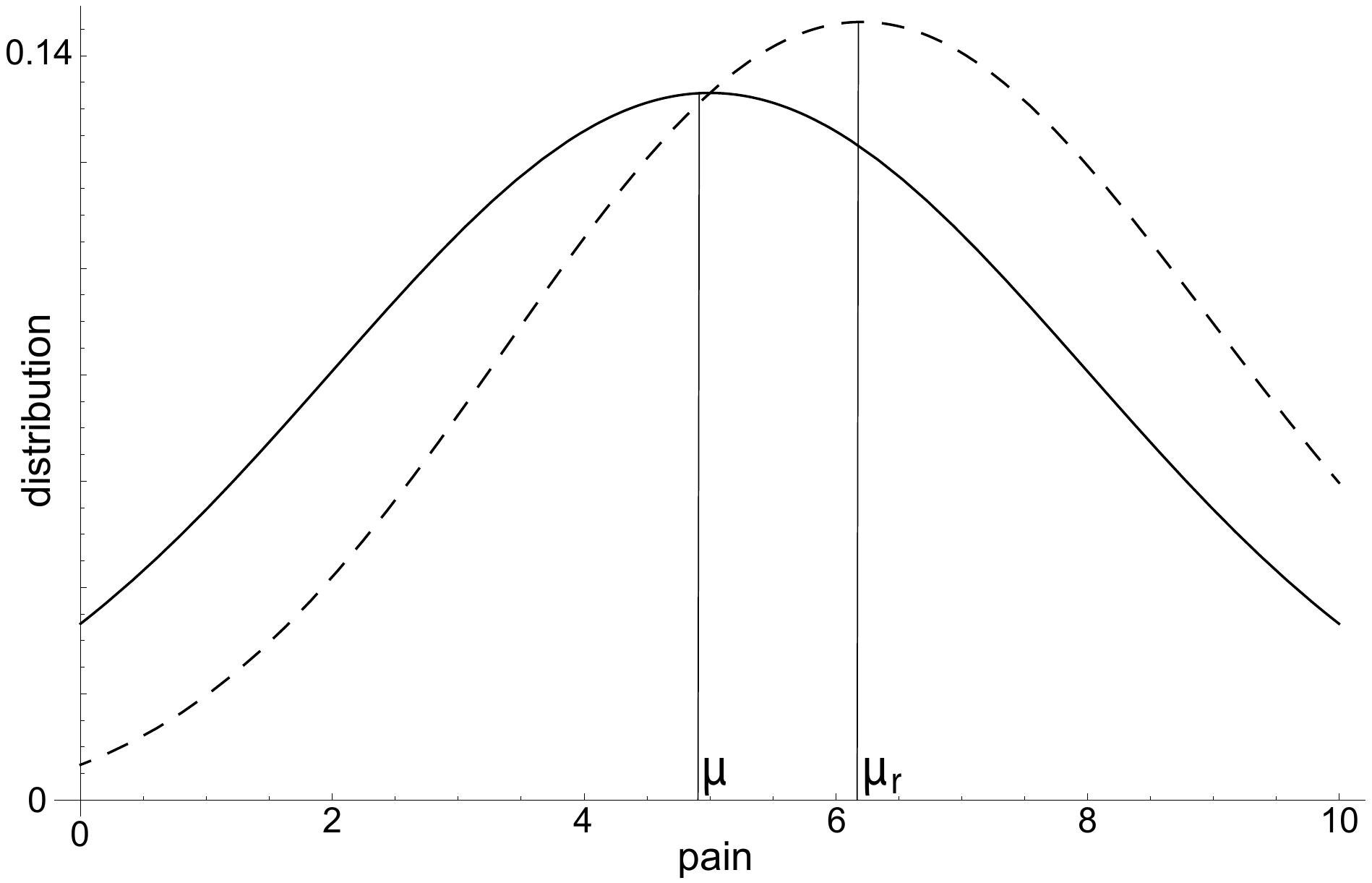}
     \caption[Probability density distributions for unbiased and biased pain reporting.]{Probability density distributions for unbiased (solid) and biased (dashed) pain reporting at a particular time $t_i$. The standard deviation (here, $\sigma=3$) has been exaggerated for illustrative purposes. In real data, the typical standard deviation is around $\sigma=1.8$.}
    \label{fig:dist}
\end{figure}

If higher pain is disproportionately reported through the mobile health application, then we will be much more likely to see higher pain levels from this normal distribution. As a first approximation, we assume the reporting bias is linear:
\begin{equation}
\rho_r (x \, | \, \mu, \sigma) = \alpha (a x + b) \exp\left[- \frac{(x-\mu)^2}{2 \sigma^2}\right] \, H(x),
\label{eq:pr}
\end{equation}
where $\alpha$ normalizes the distribution, $a, b$ tune the probabilities of reporting a pain value $x$, and the Heaviside function $H(x)$ prevents negative pain values. Figure \ref{fig:dist} shows both real and reported pain distributions at a particular time.

We need a way to connect these distributions because we want to control real pain described by \eqref{eq:p}, but the patient only provides data from the reported distribution \eqref{eq:pr}. In other words, real pain is important but invisible, and reported pain is unimportant but visible. One way to connect the distributions is through their means and variances\footnote{Note that the means and variances also change in time. We omit time dependence for notational clarity.}:
\begin{align}
\label{eq:relatep}
\mu_r (\mu,\sigma) = & \int_{-\infty}^{\infty} x \, \rho_r (x \, | \, \mu, \sigma) \, \mathrm{d}x \\
\sigma_r^2 (\mu,\sigma) = &  \int_{-\infty}^{\infty} x^2 \, \rho_r (x \, | \, \mu, \sigma) \, \mathrm{d}x.
\end{align}

Assuming $\sigma_r$ is approximately constant in time\footnote{It is not possible to verify this with data because patients only report one pain value at a particular time. However, a Kolmogorov-Smirnov normality test on residuals over the first two weeks of data rejects normality ($p < 0.05$) for only 2 of the 39 patients.}, we can also estimate $\sigma_r^2$ using the definition of variance:
\begin{equation}
\sigma_r^2 = \frac{1}{M} \sum_{i=1}^M \big(P_i - \mu_r (t_i) \big)^2,
\label{eq:vardef}
\end{equation}
where $M$ is the number of pain reports, $P_i$ is the $i$th reported pain value, and $\mu_r (t_i)$ is the expected reported pain given distribution \eqref{eq:pr} at time $t_i$.

\begin{figure}[htb!]
  \centering
    \includegraphics[width=\columnwidth]{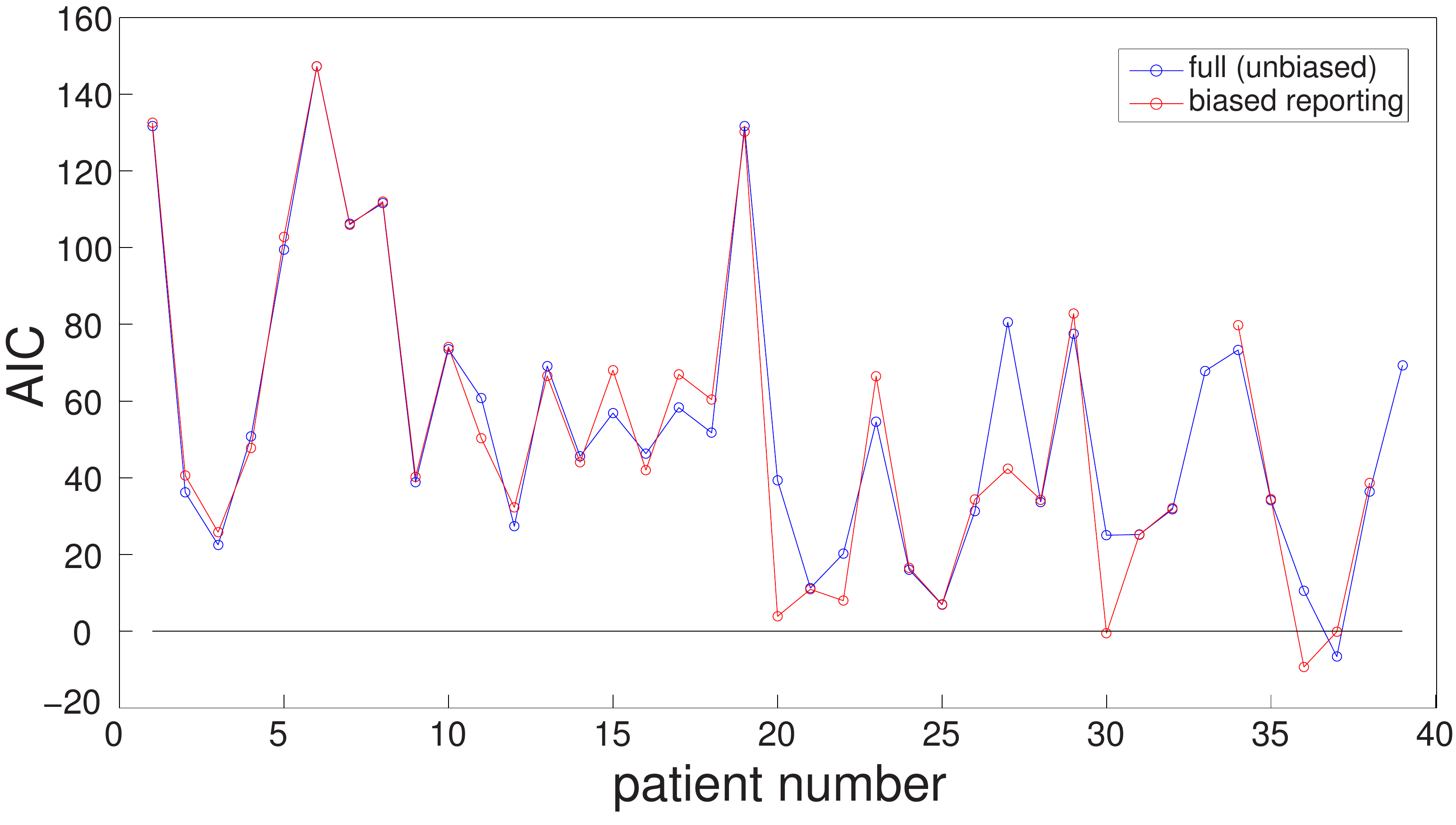}
     \caption[Akaike information criterion for unbiased and biased pain reporting.]{Akaike information criterion for unbiased and biased pain reporting. Because both models have an equal number of fitting parameters, AIC is a proxy for model likelihood (lower AIC implies higher likelihood). Again, it is not clear that one model performs universally better than the other. Note that missing biased reporting model fits indicate that the fitting algorithm did not converge ($a=0.08, b=0.1$).}
    \label{fig:aicskew}
\end{figure}

Given a proposed model for real pain, we can solve this system of three equations for the three unknowns: $\mu_r (t_i)$, $\sigma (t_i)$, and $\sigma_r$. We can then compute the likelihood of the reported pain using
\begin{equation}
\mathcal{L} = \prod_{i=1}^{M} \rho_r (P_i \, | \, \mu(t_i), \sigma(t_i)).
\label{eq:likelihood}
\end{equation}
Because we can compute the likelihood of the supplied data given any proposed model for real pain, we can tune the model parameters to maximize likelihood (technically, we minimize the negative logarithm of likelihood). This results in a best-fit model under the assumption of biased pain reporting. We compare the best-fits of the model under both biased and unbiased reporting assumptions, and find that neither model is a better fit for most patients. See Figure \ref{fig:aicskew}.

\section{Pain and medication optimization}
A key goal of the modeling of human pain dynamics is to develop predictions that allow optimized treatment: both pain and medication use should be minimized. Excess medication carries particular long-term risks for chronic pain sufferers \cite{bannwarth1999risk, gatchel2001biopsychosocial, savage1999opioid, brookoff2000chronic}, but pain mitigation is also a primary goal of SCD treatment. How can these contradictory objectives be balanced? 

Our model allows us to forecast the probability distribution of pain for a patient at a point in the near future, given past data and future drug dosage protocol. This information may be useful to a physician, allowing him or her to make an optimized, data-driven decision balancing medication and pain for the patient in real time.

\begin{figure}[htb!]
  \centering
    \includegraphics[width=0.9\columnwidth]{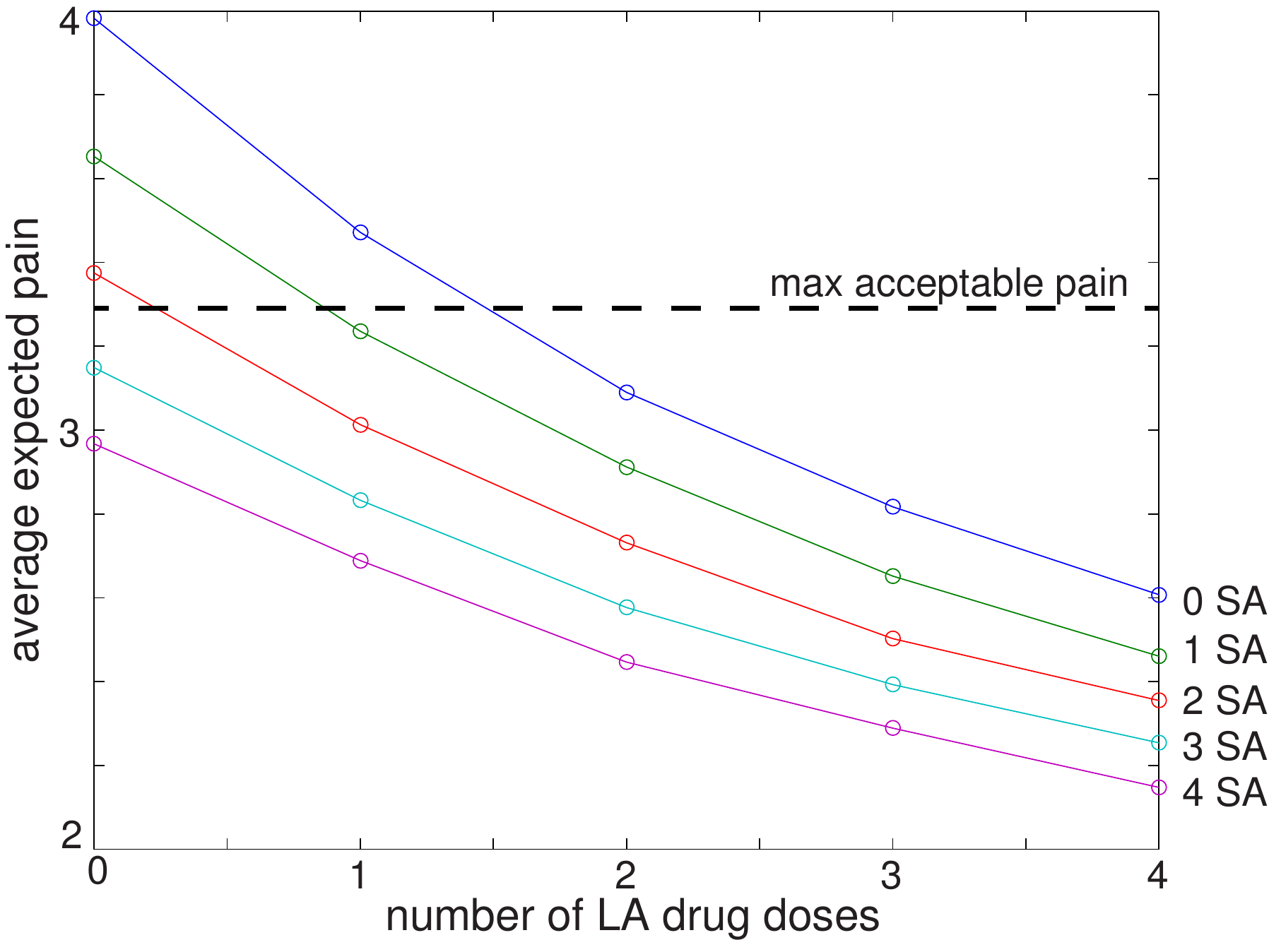}
     \caption[Example expected pain given optimal drug dosage protocol.]{Example expected pain given optimal drug dosage protocol (in this case, for Patient A3). For each set of drug dosage protocols, from no drugs to 4 doses of each drug, a physician can see the expected average pain over a certain time period. Given a patient's maximum acceptable pain level, the physician can select the best compromise between drug doses and expected pain. In this case, the physician may tell the patient to take no long-acting (LA) drugs but take 3-4 short-acting (SA) drugs. Alternatively, the physician might tell the patient to take one LA and 1-2 doses of SA medication. The timing of those LA and SA drugs is provided by the optimization algorithm.}
    \label{fig:exppain}
\end{figure}

We propose several tools that may be useful to a physician. First, we find the optimal drug timing given that a certain amount of each drug will be taken over a certain time period (say, within 24 hours). For instance, if the patient will take two short-acting opioids and one long-acting opioid within 24 hours, then the algorithm will offer the best times to take those three drug doses in order to minimize the expected average pain. We provide the physician with the expected optimal average pain for all drug combinations up to a certain maximum number of safe drug doses. See Figure \ref{fig:exppain}.

Second, we select the best drug dosing protocol (both number of drugs and dose timings) given an objective function balancing pain and medication. There are an unlimited number of possible objective functions that balance pain and medication, but we propose the following:
\begin{align} 
\label{eq:obj}
\begin{split}
m(\bar{P}, d1, d2, d3) =& \,\, w_P \Big((P_{\text{max}}-\bar{P})^{-\alpha} - P_{\text{max}}^{-\alpha} \Big) \\
                                           +& \,\, w_{d1} \Big((d1_{\text{max}}-d1)^{-\beta} - d1_{\text{max}}^{-\beta} \Big) \\
                                         +& \,\, w_{d2} \Big((d2_{\text{max}}-d2)^{-\gamma} - d2_{\text{max}}^{-\gamma} \Big) \\
                                         +& \,\, w_{d3} \Big((d3_{\text{max}}-d3)^{-\eta} - d3_{\text{max}}^{-\eta} \Big),
\end{split} \end{align}
where $\bar{P}$ is the average expected pain; $\{d1, d2, d3\}$ are the number of drug doses of each type; $\{w_P, w_{d1}, w_{d2}, w_{d3}\}$ are the weights of pain and drugs; $\{P_{\text{max}}, d1_{\text{max}}, d2_{\text{max}}, d3_{\text{max}}\}$ are the maximum safe levels of pain and drugs; and $\{ \alpha, \beta, \gamma, \eta \}$ tune the steepness of the objective function near those dangerous levels.

\begin{figure}[htb!]
  \centering
    \includegraphics[width=\columnwidth]{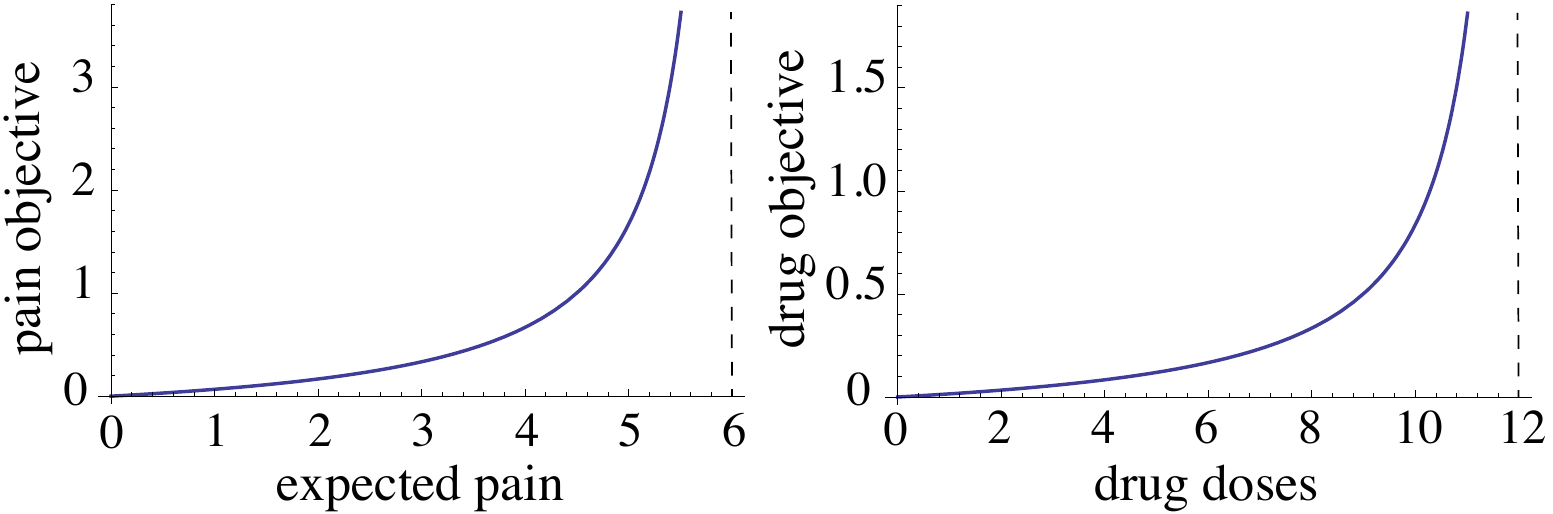}
     \caption[Contribution of the pain and drug components to the objective function.]{Contribution of the pain and one drug component to the objective function \eqref{eq:obj}. The contribution to the objective function is 0 if no pain exists or if no drugs are taken. As pain or drug doses approach dangerous levels, the contribution to the objective function blows up.}
    \label{fig:objfunc}
\end{figure}

\begin{figure}[htb!]
  \centering
    \includegraphics[width=0.9\columnwidth]{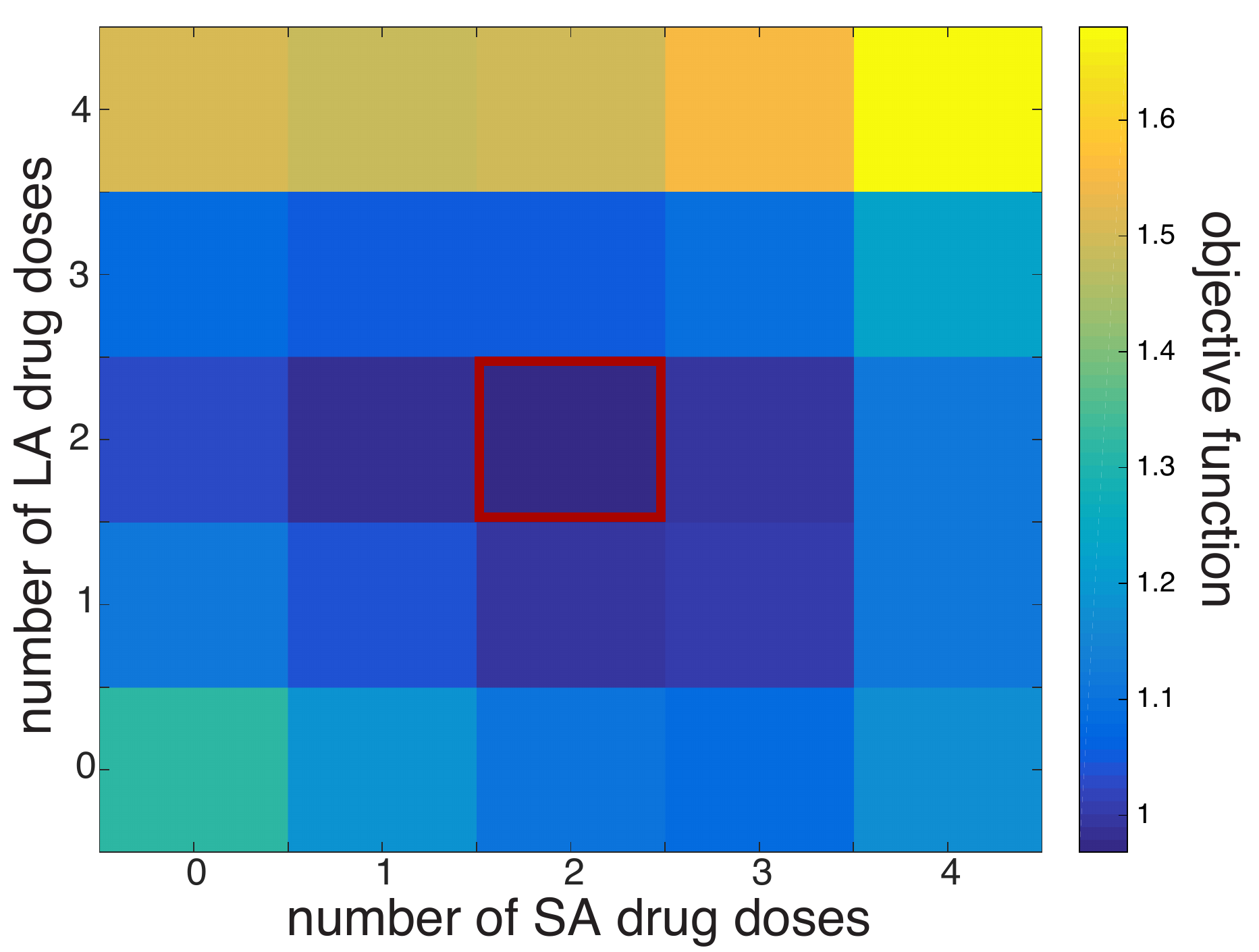}
     \caption[Example patient intervention recommendation.]{Example patient intervention recommendation (in this case, for Patient A3). For a set of personalized optimization parameters (selected by a physician or machine learning algorithm), the optimal drug timing minimizes the objective function \eqref{eq:obj} for each number of drug doses per 24-hour period. In this case, the patient is advised to take two standard doses of the long-acting (LA) opioid and two standard doses of the short-acting (SA) opioid, indicated by the red box.}
    \label{fig:objpain}
\end{figure}

See Figure \ref{fig:objfunc} for the contributions of the pain and one drug component to the objective function $m$. The contribution to the objective function is zero if no pain exists or if no drugs are taken. As pain or drug doses approach dangerous levels, the contribution to the objective function blows up. After a physician has made sufficient recommendations to a patient, a machine learning algorithm could select the weight parameters for each physician/patient pair. At that point, the algorithm could propose the optimal dosing protocol without much effort on the physician's part. See Figure \ref{fig:objpain}.

\section{Discussion and Limitations}
\subsection{Reflection on hybrid modeling}

Statistical models and mechanistic models have both been successfully applied to various aspects of human behavior.  The inference of ``black box'' statistical models from empirical data has the advantage that it obviates the need for a-priori knowledge of system dynamics.  However, mechanistic models (sometimes referred to as ``white box'' or ``clear box'') can easily incorporate such knowledge when available.

Perhaps because of the often distinct educational backgrounds of practitioners or distinct typical applications, statistical and mechanistic approaches are not frequently combined in addressing a single problem. Compared with our work, the most similar hybrid modeling idea was developed by Sheiner and colleagues in the field of pharmacokinetics, where they proposed models to estimate population characteristics of pharmacokinetic parameters \cite{sheiner1977estimation, sheiner1980evaluation, mandema1992building}. In their work, the pharmacokinetic models (i.e., mechanistic models) are well established, and the novelty and focus was the introduction of statistical models for pharmacokinetic parameter estimation. On the contrary, in our study the mechanistic model is not known before but developed by us based on clinical knowledge and reasonable assumptions, and our focus is the prediction of pain levels rather than parameter estimation.   

Other attempts based on the hybrid modeling idea in the scientific literature have appeared in the context of neural networks (e.g., \cite{psichogios1992hybrid, thompson1994modeling, su2014integrating}) and chemical engineering (e.g., \cite{thompson1994modeling, schubert1994hybrid, duarte2003hybrid}), where they largely played a computational rather than analytical role.  Some attempts have also been made with medical applications: Rosenberg et al.~(\cite{rosenberg2007using}) and  Adams et al.~(\cite{adams2007estimation}) developed a model by combining a dynamical systems approach with a statistical model to predict a patient's CD4 cell counts and HIV viral load over time in an HIV study. Timms et al.~(\cite{timms2014dynamical}) proposed a dynamical systems approach using ODEs to improve self-regulation in a smoking cessation study. Reinforcement learning techniques such as Q-learning (e.g., \cite{jaimes2014stress}) also share some commonalities with the hybrid approach. 

In this work we make our own attempt at a novel incorporation of statistical inference together with mechanistic dynamical systems modeling to produce a hybrid mathematical model for predicting and explaining human behavior.  We apply the new approach specifically to the problem of predicting the dynamics of subjective pain in a population of individuals suffering from sickle cell disease. The rationale behind our method development is that we have prior knowledge of pain trajectories with medication, making the problem suitable for mechanistic modeling; meanwhile, we do not know the relationship between patient health characteristics and pain levels, so we would like to investigate this using a statistical model.

\subsection{Limitations}
The hybrid dynamical systems/statistical approach appears to have great potential. The low frequency of pain reporting currently limits its usefulness, but future addition of high-frequency pain correlates like blood pressures, heart rate, activity level, etc., via wearable medical devices (e.g.~the ``Fitbit'') may drastically improve on that.  Furthermore, application of similar methods to more data-rich forecasting problems (e.g.~insulin levels) may also expand the utility of our work.

Another important limitation to our current model lies in the mechanistic component.  We presented here what we considered to be the simplest plausible model: pain fluctuates about an ``unmitigated'' equilibrium $u$, and medication reduces pain below that level, but pain returns as medication is metabolized and removed from the bloodstream.  This simple model cannot capture long-term changes in the unmitigated pain level, and hence its forecast validity is likely limited to short time scales (days to weeks).

Society is clearly moving in the direction of an overwhelming amount of medical data. It is imperative that we learn to take advantage of this information to improve patient treatments beyond the traditional standard of care. The approach we report here not only addresses the specific challenge of chronic pain mitigation in SCD patients, but also provides a testbed for new ways of dealing with big, ever-growing data sets in real time.

\section{Conclusions}
We have successfully demonstrated the hybrid application of statistical and mechanistic mathematical modeling with application to understanding the dynamics of subjective human pain.  Our model explains real-world data on human pain and can generate predictions of future pain dynamics.

We expect that similar methods could be used to incorporate disease-specific knowledge and modeling with statistical inference in a variety of medical applications.  Given the coming deluge of data from wearables (including clinical trial \href{https://clinicaltrials.gov/ct2/show/NCT02895841}{NCT02895841} already underway) and mobile health applications, there is a clear need for new mathematical methods to take advantage of the opportunity for personalizable, data-driven medical treatments.

%
%

\section{Data Accessibility}
Data and code available at Northwestern University ARCH repository: \url{https://doi.org/10.21985/N24S3T} \cite{NUDataSoftware}. Real patient data is not included to protect patient privacy, but all methods may be tested on synthetic data.

%
%
\section{Funding}
The authors gratefully acknowledge NSF support through grants No. DMS-1557727 and No. DGE-1324585.

%
%

\bibliographystyle{ieeetr}
\bibliography{biblio2}

\end{document}